**Main Manuscript for**

# Reconfigurable Ultrafast Thermal Metamaterial Pixel Arrays by Dual-Gate Graphene Transistors

Yibai Zhong[1], Xiu Liu[1], Zexiao Wang[1], Tianyi Huang[1], Jingyi Zou[2], Sen Lin[2], Xiao Luo[1], Zhuo Li[1], Rui Cheng[1], Xu Zhang[1,2, *], Sheng Shen[1, *]

[1]Department of Mechanical Engineering, Carnegie Mellon University, Pittsburgh, PA, USA

[2]Department of Electrical and Computer Engineering, Carnegie Mellon University, Pittsburgh, PA, USA

* Xu Zhang[1,2, *], Sheng Shen[1, *]

Email: xuzh@cmu.edu, sshen1@cmu.edu

**Author Contributions:** These authors contribute equally: Yibai Zhong, Xiu Liu, Zexiao Wang, Tianyi Huang. S.S., X.Z., and Y.Z. conceived the project with S.S. and X.Z. providing overall supervision. Y.Z., Z.W. and X.L. led the device design and optimization. T.H. led the circuit design and optimization. The electrothermal simulations were done by Y.Z. and Z.W. and the optical simulations were conducted by X.L. The graphene layers were grown and transferred by Y.Z. and J.Z., and metasurface fabrication was finished by X.L., Y.Z., and S.L., with inputs from Z.W and T.H. The experimental setup and characterization were undertaken by Y.Z., X.L., and T.H. The data analysis was led by X.L., Y.Z., S.S., and X.Z., with contributions from all co-authors including Z.L., X.Luo., R.C., J.Z., The manuscript was written by Y.Z., X.L. and S.S., with contributions from all co-authors.

**Competing Interest Statement:** S.S., X.L., and Y.Z. have filed an invention disclosure based on this work. The remaining authors declare no competing interests.

**Classification:** Physical Sciences - Applied Physical Sciences

**This PDF file includes:**

> Main Text
> Figures 1 to 4

**Abstract**

Thermal signatures represent ubiquitous infrared appearances of objects, carrying their unique spectral fingerprints. Despite extensive efforts to decipher and manipulate thermal-infrared signals, the ability to fully control them across spatial, temporal and spectral domains remains a significant challenge due to the slow speed, diffuse and broadband emitting nature of thermal emission in most materials. Here, we demonstrate a reconfigurable ultrafast thermal metamaterial pixel array that integrates active metasurfaces with dual-gate graphene transistors (Gr-FETs). The Gr-FETs with dual-gate control in each pixel achieve the heater-switch dual functionalities. As broadband transparent microheaters, Gr-FETs support the arbitrary design of integrated metasurfaces to achieve multi-color, narrowband infrared emission and operate at ultrafast modulation speeds. Concurrently as electrical switches, they enable a unified control scheme for pixel arrays of various sizes over large areas without compromising emission intensity. By decoupling the thermal generation and emission design processes, our approach provides an unprecedented degree of flexibility in programming thermal output across space, time, and wavelength. Our fabricated thermal pixel array experimentally demonstrated 26 alphabetical letters by applying progressive scanning, thus paving the way for practical realization of universal thermal signature controls for advanced thermal-infrared applications.

**Significance Statement**

Thermal signatures serve as ubiquitous spectral fingerprints for all kinds of species in nature. However, technologies for controlling and manipulating thermal signatures remain in a nascent stage due to the broadband, diffuse nature of thermal radiation and the unavoidable crosstalk between thermal signals. Here we develop a reconfigurable ultrafast thermal metamaterial pixel array by means of the monolithic integration of dual gate Gr-FETs with active infrared metasurfaces. This work establishes a new class of active thermal devices with unprecedented capabilities in space, time and frequency domains, opening avenues for adaptive thermal signaling and information-rich thermal encoding technologies.

**Main Text**

**Introduction**

As the infrared appearance of objects, thermal signatures are fundamentally governed by thermal radiation, which carries spectral fingerprints of molecular species and spans two atmospheric transparent windows[1–4]. The ubiquity and importance of thermal radiation underpin a broad range of applications, including active thermography and infrared therapy in medicine[5–9], thermal integrated photonics and active metamaterials[10–18], thermal camouflage, management and encryption[19–24], and thermal micro-electro-mechanical systems (MEMS)[25–29]. However, the ability to dynamically control thermal-infrared signatures across spatial, spectral, and temporal dimensions remains elusive due to material limitations and design complexities. While various nanophotonic structures have been employed to realize spectral, directional, and polarization control of thermal emission, these implementations are generally static and limited to single-pixel demonstrations. Recently, active metasurfaces incorporating phase-change materials, electrochemical fluids, or micromechanical machines have enabled some degree of dynamic control, but they still suffer from restricted modulation contrast, slow speed, and fabrication challenges at scale. Therefore, large-area dynamic thermal pixel arrays capable of achieving high resolution spatial modulation, high contrast and fast response are highly desired for effective manipulation of thermal signatures[30].

In the visible spectrum, the implementation of scalable pixelated arrays has been well established in visual display technologies, most notably through active-matrix electroluminescent displays utilizing thin-film transistors[31,32]. In contrast, thermal-infrared pixel techniques remain in a nascent stage. Existing work predominantly relies on direct in-plane electrical routing, which imposes constrains on pixel density, scalability and power consumption[33,34]. Recent advancements in passive-matrix configurations[35,36], wherein the pixel element is positioned at the intersection of perpendicular back column and front row electrode lines, offer partial relief by reducing wiring complexity. Yet they have been constrained by fabrication complexities and fundamentally limited by the Alt-Pleshko effect[31,37]. The realization of active thermal-infrared pixels requires devices that combine high-speed switching, broadband infrared transparency and compatibility with emitters. Graphene field-effect transistors (Gr-FETs) are uniquely positioned to meet these demands. Owing to its atomically thin nature[38–40], graphene exhibits broadband transparency consistently above 90% from visible to far-infrared wavelengths, thereby minimizing its interference with the thermal radiation signals emitted or received by the integrated devices. In addition, the ultrasmall thermal mass of graphene enables ultrafast electrothermal responses[41–44]. Nevertheless, conventional Gr-FETs suffer from substantial leakage currents due to monolayer graphene (MLG)'s semi-metallic nature[45–47], resulting in significant emission crosstalk [48].

In this work, we develop a reconfigurable thermal metamaterial pixel architecture that overcomes these longstanding barriers through the monolithic integration of dual gate Gr-FETs with active infrared metasurfaces. The dual-gate control of each pixel enables "heater-switch" duality for Gr-FETs, where transistors can not only behave as transparent tunable microheaters integrated with arbitrary plasmonic metamaterials for thermal emission control, but also act as analog switches for thermal power regulation, simultaneously maximizing thermal contrast and minimizing inter-pixel emission crosstalk. The dual-gate design enables dynamic tuning of thermal-infrared pixel power distribution and thermal emission, without the need for complex voltage routing or mechanical modulation. Our transient measurement further verifies that the switching speed of such thermal pixels can be ultrafast owing to extremely small thermal mass of MLG, without compromising its emission intensity. Moreover, since large-area MLG can be synthesized via cost-effective low pressure chemical vapor deposition (LPCVD) production with simple carbon precursor[49–52] and transferred onto almost all substrates for van der Waals integration[53–55], we experimentally realize a large-area active matrix thermal pixel array operating under a unified gate-driving scheme and demonstrate programmable emission patterns, including full Latin alphabet rendering, through progressive scanning. By decoupling the heat control from thermal-infrared emission shaping at the pixel level, this platform introduces a scalable strategy for spatiotemporal and spectral control of infrared signals with sub-millisecond response, paving the way for advanced thermal displays, adaptive camouflage and high-speed infrared communication.

**Results and Discussion**

The core of our system is a reconfigurable thermal metamaterial pixel, designed to deliver localized, spectrally selective, and rapidly switchable infrared emission. Each pixel consists of a central metasurface-integrated microscale graphene thermal engine (MM Gr-FET) surrounded by four peripheral-unit graphene transistors (PU Gr-FETs) configured as voltage-controlled switches. Figure 1a shows the design concept of the thermal metamaterial pixel array. To allow for scalable integration, we arrange pixels in rows and columns with shared source-drain and gate lines, implementing the active-matrix addressing schemes used in visual display technology. Within each pixel, the MM Gr-FET acts as the primary emitter, delivering Joule heating via source ($V_S$) and drain ($V_D$) to a spectrally engineered metasurface that defines the emission wavelength. The PU Gr-FETs act as tunable resistive pathways that gate power delivery to the MM device, enabling a high degree of thermal contrast through selective voltage gating. The bottom gates and the source/drain lines for all transistors are fabricated of Au with aluminum oxide ($Al_2O_3$) as the dielectric layer and thermal oxide ($SiO_2$) Si wafer as the substrate. All the fabrication steps are compatible with the complementary metal-oxide-semiconductor (CMOS) process, as discussed in Methods. As demonstrated in the circuit diagram Fig. 1b, the four PU transistors parallelly connect the $V_D$ of the pixel on one side and the MM transistor on the other side, which then connects the $V_S$ of the pixel via the MM transistor, as demonstrated in the false-color SEM image (Fig. 1c). Since $V_D > V_S$ is commonly set for the pixel, the four PU Gr-FETs can effectively pull up the potential across the MM transistor when their channel resistances are lowered by the shared bottom gates ($V_{G,PU}$). Meanwhile, the MM transistor controlled by another bottom gate $V_{G,MM}$ can have their channel resistance varied simultaneously to increase or decrease their surface temperature rise. Instead of allocating individual gates for each pixel, the pixels along the same y-direction share the same gate lines, therefore significantly reduces the control complexity. Yet, each pixel is individually addressable when power inputs are applied across one designated row, crossing the two voltage potentials applied along one pixel column.

Such a configuration allows pixel-level switching without altering the source-drain bias, dramatically reducing circuit complexity and power overhead as arrays scale. Meanwhile, since thermal emission results from the product of emissivity and Planck blackbody radiation, we strategically design the pixel region (Fig. 1d) with metamaterials exhibiting near-unit emissivity resonance, while the surrounding PU regions remain spectrally neutral due to graphene's low emissivity and infrared transparency. This configuration enhances emission contrast, complementing the temperature difference achieved through our pixel circuit architecture. The metamaterials are also highly customizable – the geometry variation of their units would lead to narrow-band thermal emission peaking at different wavelengths, as demonstrated in Fig. 1e. By separating the mechanisms of heat generation and spectral emission, this pixel architecture enables unmatched flexibility in dynamically programming thermal output across spatial, temporal, and spectral dimensions.

To evaluate the electrothermal performance of the pixel array, we first characterize the electrical and material properties of the monolayer graphene synthesized via LPCVD (details in Methods). Gr-FET test structures, as shown in Figs. 2a-inset, are fabricated following the same fabrication steps as the pixelated devices (details in Methods) to assess their carrier mobility and contact resistance. Its channel width is 20 μm with four different channel lengths of 5 μm, 10 μm, 20 μm and 50 μm so that the contact resistance and mobility of MLG can be measured via transmission line method (TLM)[56]. When a constant source-drain voltage $V_{DS}$ = 50 mV is applied across each of the channels, the contact resistance of the device is estimated to be around 200 Ω, (detailed in Supplementary Note 1), which is more than one order of magnitude smaller than that of the graphene film and hence can be neglected. The field-effect carrier mobility of graphene $\mu$ can be modelled using the Drude model with the $I_{DS}$-$V_G$ measurement results shown in Fig. 2a:

$$\mu = \frac{L}{W}\frac{1}{C'_{OX}}\frac{1}{V_{DS}}\frac{dI_{DS}}{dV_G} , \qquad (1)$$

where $L$ and $W$ are channel length and width of the Gr-FET, $I_{DS}$ and $V_G$ are the measured channel current and gate voltage set for the Gr-FET, and $C'_{OX} = \varepsilon_r \varepsilon_0 / d$ representing the capacitance of Gr-FET per unit area with $\varepsilon_r$ equaling the relative permittivity of $Al_2O_3$ dielectric layer and $\varepsilon_0$ being the vacuum permittivity. The maximum hole mobility is estimated to be around 1100 cm$^2$/V·s, with the Dirac point being close to neutral gate voltage, which indicates a good carrier mobility with minimum doping for LPCVD-synthesized graphene. Raman spectroscopy confirms monolayer graphene's quality and low defect density (Supplementary Note 2), validating that the LPCVD graphene is suitable for high performance thermal modulation. Based on high quality MLG, we then design and fabricate thermal metamaterial pixels, as shown in Fig. 1c. Each pixel consists of a central square-shaped MM Gr-FET with a side length of 20 μm covered by Au metasurface layer, which serves as an emitter. Surrounding the MM Gr-FET are four L-shaped PU Gr-FETs, left optically transparent to minimize background emission. The four PU Gr-FETs, when connected in parallel, are designed with a larger surface area than the MM transistor to dissipate residual heat. Under identical gate voltage, the PU Gr-FETs exhibit nearly four times higher resistance than the MM Gr-FET. The electrical conductivities of the two sets of Gr-FETs ($\sigma_{MM}$ and $\sigma_{PU}$) are governed by both field-induced charge carriers (tunable via gate bias) and residual carriers ($n_{res}$) accounting for charge transport near the Dirac point[45,47]. These conductivities can be modelled as:

$$\begin{cases} \sigma_{MM} = n_{res}e\mu + C'_{ox}|V_{G,MM} - V_{Dirac,MM}|\mu \\ \sigma_{PU} = n_{res}e\mu + C'_{ox}|V_{G,PU} - V_{Dirac,PU}|\mu \end{cases}, \quad (2)$$

where $V_{G,MM} - V_{Dirac,MM}$ and $V_{G,PU} - V_{Dirac,PU}$ are the potential difference between gate voltages and Dirac points of MM Gr-FET and PU Gr-FET, respectively. The $n_{res}e\mu$ term captures the residual conductivity of graphene at the Dirac point, which arises from diffusive transport driven by carrier density fluctuations, induced by charged impurities typically located in the substrate or at the graphene-substrate interface [57]. Assuming negligible contact resistance, the total channel resistance of the thermal metamaterial pixel can hence be expressed as:

$$R_{channel} = \frac{1}{\sigma_{MM}} \frac{L_{MM}}{W_{MM}} + \frac{1}{\sigma_{PU}} \frac{L_{PU}}{W_{PU}}, \quad (3)$$

where $\frac{1}{\sigma_{MM}} \frac{L_{MM}}{W_{MM}} = R_{MM}$, $\frac{1}{\sigma_{PU}} \frac{L_{PU}}{W_{PU}} = R_{PU}$, accounting for the contributions from the MM Gr-FET and the PU Gr-FETs, respectively. The effective geometric aspect ratio $\frac{L_{MM}}{W_{MM}} = 1$ and $\frac{L_{PU}}{W_{PU}} = 3.91$. Therefore, the thermal metamaterial pixel can be turned on and off via concurrent gate tuning of both the MM Gr-FET ($V_{G,MM}$) and the PU Gr-FETs ($V_{G,PU}$). As measured in Figs. 2b, when $V_{DS}$ = 50 mV and $V_{G,PU}$ = 3 V (set close to the Dirac point of the PU Gr-FETs), the overall $I_{DS}$ of the pixel is small regardless of $V_{G,MM}$ due to the dominant larger resistance of PU Gr-FETs. Further biasing $V_{G,MM}$ negatively reduces the MM Gr-FET channel resistance, resulting in the majority of the voltage drop to occur across the PU Gr-FETs, minimizing the power dissipation in the MM Gr-FET microheater and effectively turning the pixel off, as evidenced by the measured $I_{DS}$ in Fig. 2b using Eqns. 2 and 3 and demonstrated in Fig. 2c (detailed calculation in Supplementary Note 3). On the contrary, when $V_{G,MM}$ = 4 V (near the Dirac point of the MM Gr-FET) and when $V_{G,PU}$ is set negatively away from the Dirac point, the pixel can be turned on and the power is funneled into the MM Gr-FET microheater. This configuration results in an increase in the total $I_{DS}$ compared with the off case (Fig. 2b), and localizes heating at the emitter, as shown in Fig. 2c.

To model and optimize the pixel's thermal response under dual-gate control, we establish a steady state thermoelectrical simulation via COMSOL Multiphysics. The source-drain voltage $V_{DS}$ is set to a constant 7 V, while the background temperature is set to 298 K. The model reveals strong thermal localization in the MM Gr-FET under appropriate gate biasing. The pixel is turned on when $V_{G,MM}$ = 7.9 V and $V_{G,PU}$ = -4.4 V, as shown in Fig. 2d, where the MM Gr-FET reaches a temperature rise of nearly 15 K while the PU Gr-FETs have a lower temperature rise of 3 K. On the other hand, when $V_{G,MM}$ = -8.6 V and $V_{G,PU}$ = 4.2 V, the pixel is turned off as displayed in Fig. 2e, where temperature rise at the MM Gr-FET is 2 K while the PU Gr-FETs maintain a temperature rise of around 3 K. Such predicted thermal performance is further verified by thermal mapping under an infrared camera shown in Figs. 2f and 2g. Notably, the residual heat from the PU Gr-FETs is consistently invisible under thermal mapping due to the high transparency of monolayer graphene and the low emissivity of Au underneath. Meanwhile, > 15 K temperature rise from the MM Gr-FET is captured when the pixel is on, while its temperature rise is < 3 K when the pixel is turned off, demonstrating high spatial thermal contrast controlled purely by electrostatic gating. The lack of visible emission from the PU Gr-FETs further confirms the high IR transparency of monolayer graphene and the spectral selectivity of the metasurface design. This new methodology paves the way for implementing large scale infrared pixel arrays without compromising control capability or thermal contrast. To assess the switching speed and cutoff frequencies of the thermal metamaterial pixels, time-domain thermoreflectance measurements are conducted with respect to the MM transistor. As plotted in Fig. 3a, by alternating the gate voltages between the on-state pair ($V_{G,MM}$ = 3 V and $V_{G,PU}$ = -6 V) and off-state pair ($V_{G,MM}$ = -9 V and $V_{G,PU}$ = 3 V) every 7.5 μs, with $V_{DS}$ is maintained at 5 V, we measure the temporal evolution of pixel heating and cooling. The simulated temperature rise and the measured reflectance change exhibit excellent agreement in their dynamics. When turned on and off, transient simulations show the 10-90% rising time $t_r$ and falling time $t_f$ of 2.73 μs and 2.77 μs, respectively, corresponding to a minimum 3-dB cutoff frequency of $f_c$=0.35/$t_f$=126 kHz [48]. As captured in Figs. 3b to 3e (image capturing detailed in Methods), the reflectance change $\Delta R/R$ captured from the heater surface is found to decrease when the pixel switches from off to on state and increase back to zero when the pixel turns off (Supplementary Video), with the 10-90% rising time being 1.87 μs and the falling time being 1.33 μs, corresponding to a minimum 3-dB cutoff frequency of 187 kHz. These results confirm that such pixel switching is governed by the ultralow thermal mass of monolayer graphene, which is a key advantage compared with the slower timescales typical of electrochemical materials or bulk heaters[58–61].

To demonstrate the scalability and programmability of our platform, we fabricate an array of thermal metamaterial pixels as a reconfigurable infrared display. The fabricated device is wire bonded onto a chip holder (Fig. 4a) and integrated into external controlling circuits. As displayed in Fig. 4b, the thermal metamaterial array consists of 9 pixels arranged into three rows by three columns. The parallel connection of pixels in each row enables consistent source/drain voltage configurations across each pixel, enabling active-matrix style addressing with minimal wiring complexity. As a proof of concept, we implement a progressive scanning protocol that activates specific pixels in sequence to form programmable infrared patterns. To demonstrate the ability of individually turning on the center pixel A (dashed box in Fig. 4b) and keeping the remaining ones off, $V_{DS}$ = 5 V ($V_D$ = 5 V and $V_S$ = 0 V) is set for the middle row where pixel A is located. For the row above pixel A, all the source/drain lines share the same voltage as $V_S$, and for the row below pixel A, $V_D$ is set for the remaining source/drain lines. Therefore, no other pixel rows are illuminated because of the net zero voltage drop. $V_{G,MM}$ = 3 V and $V_{G,PU}$ = -6 V are hence applied to the gate columns for pixel A to turn it on, while $V_{G,MM}$ = -9 V and $V_{G,PU}$ = 6 V are maintained for the other columns to mute the remaining pixels on the same row. Similar procedures can be followed for any other pixels, which are demonstrated in Supplementary Note 4. Meanwhile, one can also set $V_{G,MM}$ = 3 V and $V_{G,PU}$ = -6 V for multiple gate voltage columns, effectively turning on many or even all the pixels on the same row simultaneously without compromising their intensities (details shown in Supplementary Note 4). Such flexible control schemes combined with fast transient response of pixels enable progressive scanning for the entire thermal metamaterial pixel array to form dynamic patterns. As demonstrated in Fig. 4c, the 9-pixel array scans through each row of pixels with one, two or three pixels light up at the desired locations. Using this approach, we successfully render all 26 Latin alphabet letters in real-time by scanning through pixel combinations across the array. This demonstration demonstrates the feasibility of using our architecture for reconfigurable thermal displays and dynamic thermal encoding at the microscale.

**Conclusion**

In summary, we demonstrate a CMOS-compatible platform for reconfigurable pixelated thermal metamaterial arrays that achieve ultrafast, high contrast and spectrally selective infrared emission via electrostatic gating of dual-function Gr-FETs. With three-by-three thermal metamaterial pixel arrays, we demonstrate the display of all the 26 Latin alphabet letters via progressive scanning. The ultralow emissivity of MLG decouples temperature and emission patterns and enables flexible integration of various thermally excited infrared emitting materials such as metasurfaces and nanoparticles with diverse materials and geometries, thus allowing highly customizable spectrum, polarization, and direction of the infrared emission. Our demonstration of sub-millisecond switching, narrowband spectral tunability, and programmable 2D pattern formation establishes a new class of active thermal devices with unique capabilities in space, time, and frequency domains, thus addressing the increasing demand of advanced infrared and thermal applications at micro and nanoscale.

## Materials and Methods

Graphene Synthesis

The MLG is synthesized via LPCVD using a 2-inch-diameter quartz tube furnace. The substrate is 99.9% pure Cu foil (25 µm thick) which is cut into one-inch-square pieces and polished via electrochemical polishing for 2.5 hours in 85% phosphoric acid as electrolyte. Ultra-high purity hydrogen with flow rate of 70 standard cubic centimeters per minute (sccm) is purged into the LPCVD chamber when the furnace is heated up to 1000 ºC in 15 minutes. The temperature and hydrogen flow are maintained for another 30 minutes to further anneal the Cu foil. Then the ultra-high purity methane gas acting as the carbon precursor is mixed into the chamber with hydrogen: methane = 65 sccm: 5 sccm to maintain the total flow rate for another 30 minutes. Methane molecules hence decompose into a mixture of hydrogen, hydrocarbon and carbon with Cu as catalyst, leaving MLG deposited onto the Cu substrate. After the MLG deposition is finished, the methane flow is turned off and the furnace is opened to allow a rapid cooling process, with a flow of hydrogen of 200 sccm until room temperature is reached.

Device Fabrication

The Gr-FET is fabricated on a Si wafer with 1 µm thick thermal oxide. Photolithography followed by e-beam evaporation of 50 nm thick Au is used to fabricate the bottom gate electrode with metal lift-off. Thermal atomic-layer deposition (ALD) is used to deposit an $Al_2O_3$ dielectric layer of 40 nm thickness. The LPCVD-synthesized MLG is wet transferred to the top of the dielectric layer with PMMA as supporting layer and patterned to desired shape via the same photolithography process and reactive-ion etching (RIE). Lastly, similar metal deposition procedures are followed to form the source and drain electrodes. E-beam lithography is used for defining Au metasurfaces to enhance infrared emission, followed by e-beam evaporation (50 nm Au) and metal lift-off processes.

Thermal Mapping and Thermoreflectance Measurements

A QFI InfraScope is employed to capture dynamic thermal mapping images. The thermal pixels are coated with a layer of Au metasurfaces exhibiting an emissivity close to unity at wavelength of 2.9 µm, ensuring compatibility with the 2.0-4.5 µm response range of the thermal mapping system. The thermoreflectance measurement employs a 530-nm laser, where Au emitter is witnessed to have the largest negative reflectance-temperature gradient under such incident wavelength[62–64]. Two arbitrary-wave generators are synchronized by a master-slave trigger cable in finite-burst mode to simultaneously modulate the two gate voltages following a square wave (15 µs period; 50% duty cycle). The movie mode is used to record and extract all captured frames during the modulation process.

## Acknowledgments

This work is funded by the Defense Threat Reduction Agency (Grant No. HDTRA1-19-1-0028), National Science Foundation (Grant No. ECCS-2426252), National Science Foundation (Grant No. ECCS−2239822), and National Science Foundation (Grant No. CBET-1931964).

**Figures and Tables**

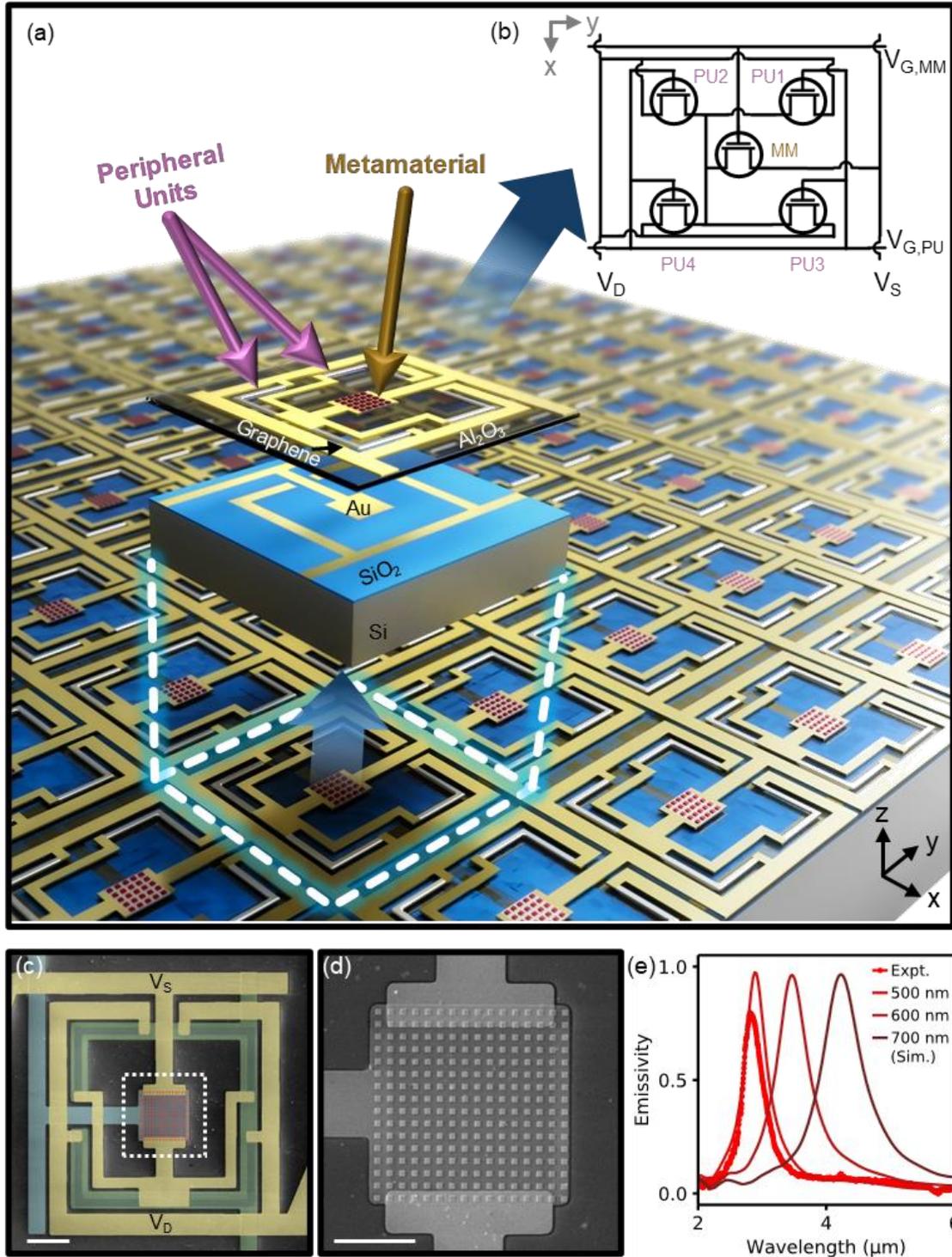

**Figure 1.** Reconfigurable ultrafast thermal metamaterial pixel array. (a) Schematic of the active-matrix thermal metamaterial pixel architecture, composed of pixel rows (x-direction) with parallel connectivity and shared control gates along the y-direction, enabling a unified gate-driving scheme. The zoom-in view of one thermal metamaterial pixel unit consists of an MM Gr-FET located in the center of the pixel and four L-shaped PU Gr-FETs surrounding it, with two gate

electrodes controlling both. (b) Circuit diagram of one thermal metamaterial pixel unit. Four analog voltage inputs are needed to operate one pixel: The four PU Gr-FETs are connected in parallel with each other and are controlled by the same gate ($V_{G,PU}$). The MM Gr-FET is controlled by another gate ($V_{G,MM}$) and connects in series with the PU Gr-FETs. Each pixel on the same row shares the same $V_S$ and $V_D$. (c) Scanning electron microscopy (SEM) image with false colors of a thermal metamaterial pixel unit (scale bar: 20 μm). The source/drain lines are colored yellow, and the graphene area are colored blue. The cyan and green colors represent gate electrodes for the MM Gr-FET and PU Gr-FETs. (d) Zoomed-in view showing the metamaterial on top of the MM Gr-FET (scale bar:10 μm). (e) Emissivity simulation of Au metasurfaces with side length of 500 nm, 600 nm and 700 nm. The experimental measurement of a fabricated metasurface with side length of 500 nm shows good agreement with the simulation.

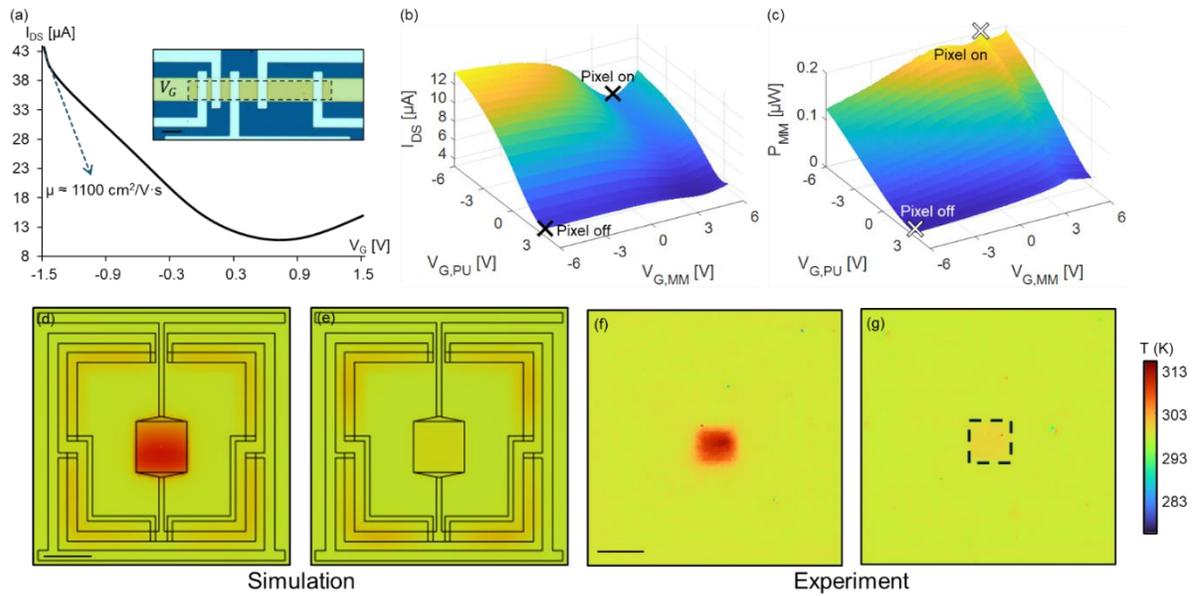

**Figure 2.** Characterization of the MLG and single thermal metamaterial pixels. (a) Electrical $I_{DS}$-$V_G$ characterization with $V_{DS}$ = 50 mV on graphene FET test structures (inset with scale bar: 20 µm). The maximum charge carrier mobility estimated from the curve gradient is 1100 cm$^2$/V·s. (b) Dual-gate electrical characterization of the thermal metamaterial pixel. The $I_{DS}$-$V_G$ measurement shows that the overall channel resistance of the pixel increases as $V_{G,PU}$ changes from -6 V to 3 V. The variation of $V_{G,MM}$ further changes the distribution of internal power, turning the pixel off by decreasing the channel resistance of the heater transistor when the overall channel resistance is high ("Pixel off"), and vice versa ("Pixel on"). (c) Calculated power distribution on the MM Gr-FET verifies the aforementioned configurations. (d) to (g) Steady-state simulation and experiment thermal characterization of the pixel (scale bar: 20 µm). $V_{DS}$ = 7 V. $V_{G,MM}$ = 7.9 V, $V_{G,PU}$ = -4.4 V when pixel-on, $V_{G,MM}$ = -8.6 V, $V_{G,PU}$ = 4.2 V when pixel-off.

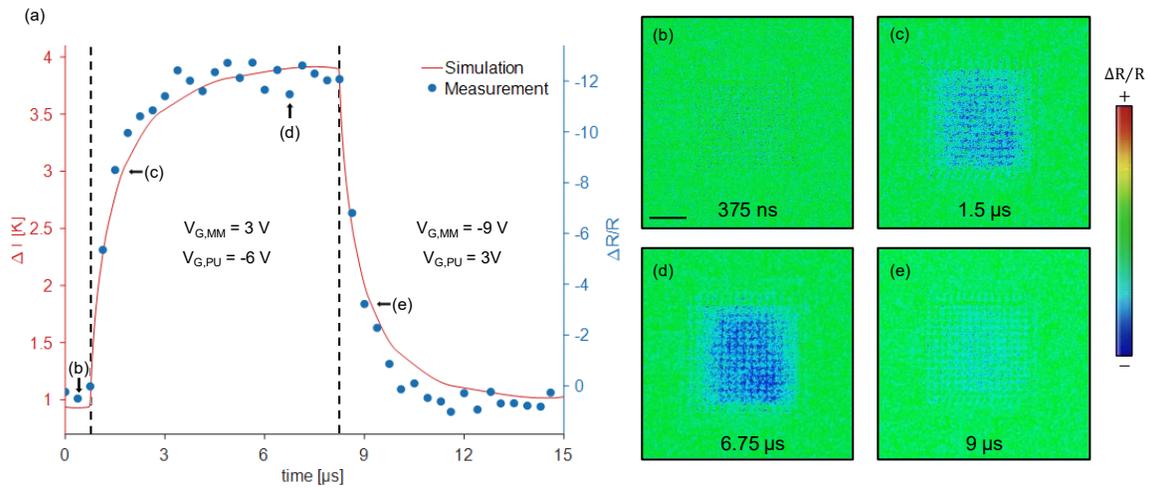

**Figure 3.** Transient analysis of single thermal metamaterial pixels. (a) Temperature simulation and thermoreflectance measurement of the MM Gr-FET area under a square-wave gate voltage modulation (15 μs period; 50% duty cycle), with key frames of interest captured in (b) to (e). $V_{DS}$ is maintained at constant 5 V for both analyses. The minimum 3-dB cutoff frequencies are calculated to be 126 kHz for the simulation and 187 kHz under measurement.

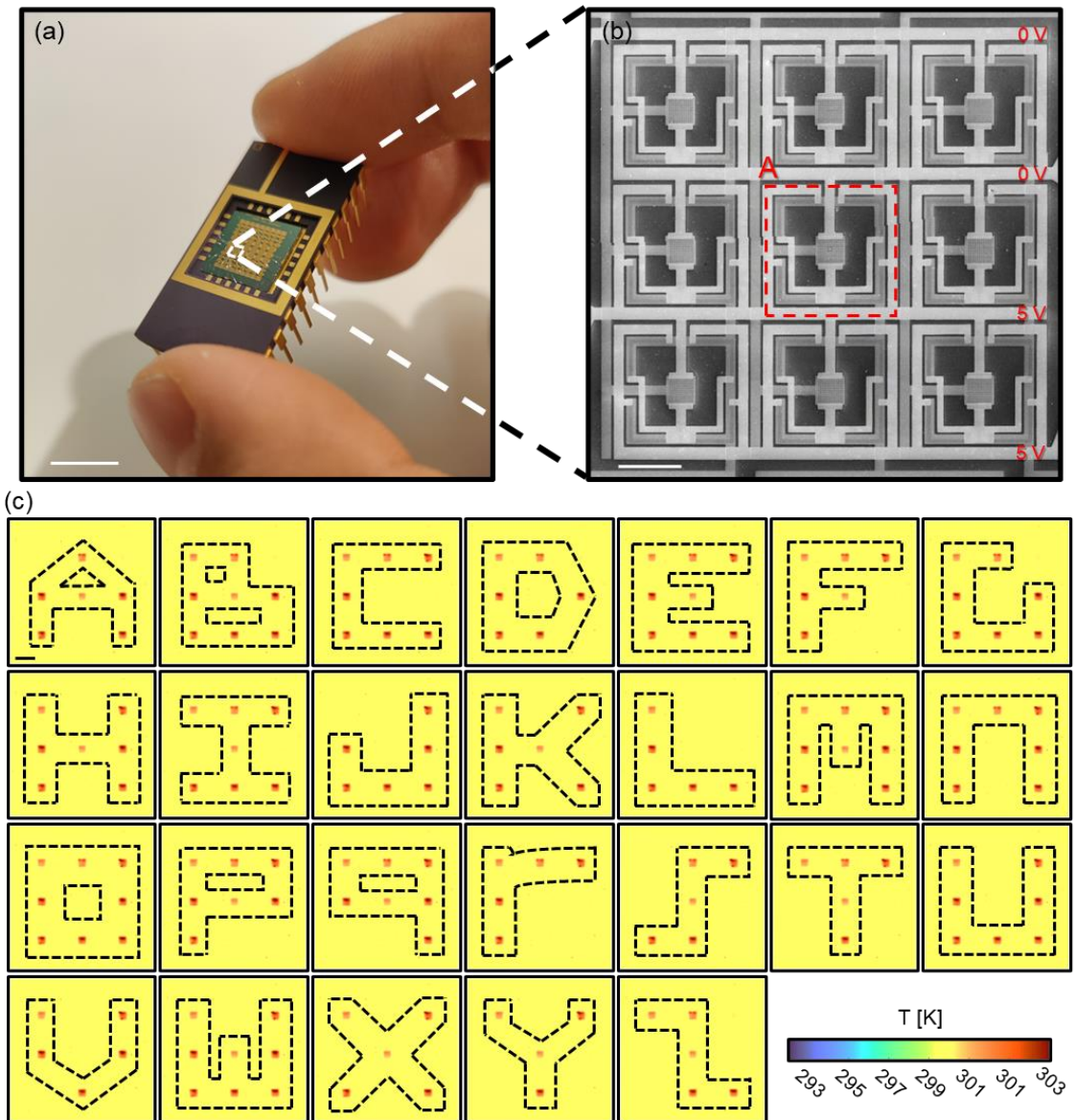

**Figure 4.** Reconfigurable three-by-three thermal metamaterial pixel array. (a) Wire-bonded devices onto a 24-pin chip holder containing multiple three-by-three thermal metamaterial pixel arrays (scale bar: 90 mm). (b) Zoomed-in SEM image of a three-by-three thermal metamaterial pixel array (scale bar: 50 μm). (c) Progressive scan across the array displaying the complete Latin alphabets (scale bar: 50 μm). The 5 V potential difference created between one selected pair of source/drain lines enables individual pixel or full-row addressing with tuning of gate voltages.

16